\documentclass[aps,twocolumn,superscriptaddress,amsfont,amssymb,amsmath,showpacs,balancelastpage,english]{revtex4-1} 

\usepackage[dvipdfmx]{graphicx}
\usepackage{dcolumn}
\usepackage{enumerate}
\usepackage{bm}
\usepackage{comment}
\usepackage{url}
\usepackage{color}
\usepackage{tikz}
\usepackage{mathtools}

\usepackage[utf8]{inputenc}
\usepackage[T1]{fontenc}

\newcommand{\eq}[2][]{
	\begin{align#1}
		#2
	\end{align#1}
}

\newcommand{\diff}{\mathrm{d}}
\newcommand{\iu}{\mathrm{i}}
\newcommand{\e}{\mathrm{e}}
\newcommand{\sinc}{{\rm{sinc}}}

\newcommand{\figref}[1]{Fig.~\ref{#1}}
\newcommand{\tabref}[1]{Table~\ref{#1}}
\newcommand{\secref}[1]{Sec.~\ref{#1}}
\newcommand{\appref}[1]{Appendix~\ref{#1}}

\newcommand{\typicalparam}[2]{\left( \frac{#1}{#2} \right)}

\allowdisplaybreaks[4] 

\definecolor{mygreen}{RGB}{0,130,0}

\begin{document}


\title{Observational constraint on axion dark matter in a realistic halo profile with gravitational waves}

\author{Takuya Tsutsui}
\affiliation{%
Research Center for the Early Universe (RESCEU), Graduate School of Science, The University of Tokyo, Tokyo 113-0033, Japan
}
\affiliation{%
Department of Physics, Graduate School of Science, The University of Tokyo, Tokyo 113-0033, Japan
}
\author{Atsushi Nishizawa}%
\affiliation{%
Research Center for the Early Universe (RESCEU), Graduate School of Science, The University of Tokyo, Tokyo 113-0033, Japan
}
\email{anishi@resceu.s.u-tokyo.ac.jp}

\date{\today}

\begin{abstract}
Axions are considered as a potential candidate of dark matter. The axions form coherent clouds, which can delay and amplify gravitational waves (GWs) at a resonant frequency and produce secondary GWs following a primary wave. All GWs detected so far from compact binary mergers propagate in the Milky Way halo, which is composed of axion dark matter clouds, resulting in the production of the secondary GWs. The properties of these secondary GWs depend on the axion mass and its coupling with the parity-violating sector of gravity.
In our previous study, we have developed a search method optimized for the axion signals and obtained a constraint on the coupling that was approximately ten times stronger than the previous best constraint for the axion mass range, $[1.7 \times 10^{-13},\,8.5 \times 10^{-12}]\,\mathrm{eV}$. However, the previous search assumed that dark matter was homogeneously distributed in the Milky Way halo, which is not realistic.
In this paper, we extend the dark matter profile to a more realistic one called the core NFW profile. The results show that the previous study's constraint on the axion coupling is robust to the dark matter profile.

\end{abstract}

                            
\maketitle



\section{Introduction} \label{sec:introduction}
The $\Lambda$ cold dark matter (DM) model is currently the most successful cosmological model. It assumes that the Universe is homogeneous and isotropic and contains dark energy, baryons, and cold DM. Using data from the baryon acoustic oscillations of galaxy clustering observed by the Sloan Digital Sky Survey III, and the cosmic microwave backgrounds observed by WMAP and Planck, the cosmological ingredients of energy at present have been estimated to be the dark energy of $\sim 68\%$, the cold DM of $\sim 27\%$, and the baryon of $\sim 5\%$, e.g.,~\cite{baryon_oscillation_SDSS}.

There have been many candidates considered for dark matter, including axions. Historically, axions were introduced as a solution to the strong-CP problem. although the quantum chromodynamics (QCD) Lagrangian has a CP-violating term that make a vacuum gauge-invariant~\cite{review_CP_problem}, many experiments support the conservation of the CP symmetry. (e.g. from the measurements of the electric dipole moment of neutrons~\cite{electric_dipole_moment}, the violation of the CP symmetry must be suppressed at the level of $\sim 10^{-10}$). The measurements of the electric dipole moment of neutrons indicate that the violation of CP symmetry must be suppressed to the level of $\sim 10^{-10}$. To solve this problem, Peccei and Quinn introduced a new global symmetry of chiral $U(1)$ anomaly~\cite{peccei_quinn}, called the PQ symmetry. With the spontaneous breaking of this symmetry, axions appear as a pseudo-Nambu-Goldstone boson. The axion Lagrangian also has a CP-violating coupling term between the axion field and the gauge field (the gluon field). As a result of the spontaneous symmetry breaking, the axion field is dynamically adjusted to cancel out the CP-violating term in the QCD Lagrangian, solving the strong-CP problem.

The axions introduced by Peccei and Quinn are known as QCD axions. By considering interactions between the QCD axions and neutral pions, the axion mass can be related to the axion and pion decay constants, as well as the pion mass. The cosmological evolution of axions has been studied in~\cite{cosmological_axion1, cosmological_axion2, cosmological_axion3, cosmological_axion8, cosmological_axion6, cosmological_axion7}, while recent developments in search methods and observational constraints are reviewed in~\cite{review_axion_EM, DiLuzio:2020wdo, Galanti:2022ijh}. There are other types of axions known as axion-like particles, which are expected to exist in string theory~\cite{string_axiverse}. The string theory predicts the existence of the extra dimensions~\cite{becker_becker, zwiebach, polchinski}, and due to the topological complexity of the extra-dimensional manifold, the mass range of the axion-like particles can vary widely. Therefore, this paper does not limit the search to specific types of axions, but rather investigates a broad mass range of axion-like particles.

Axions are massive and can couple to gravity. Since a gauge field is described by the Riemann tensor in general relativity (GR), we consider the Chern-Simons (CS) interaction, that is, the axion-CS gravity~\cite{review_CSgravity}. This is a low-energy effective theory of the parity-violating extension of GR. In the axion-CS gravity, GWs stimulate axion decay into gravitons and are enhanced during propagation~\cite{soda-yoshida, soda-urakawa}.

In this paper, we consider the Milky Way (MW) DM halo composed of axions with the gravitational coupling mentioned above. In the model, due to the axion decay, all GWs propagating in the MW DM halo are amplified and delayed at the resonant frequency corresponding to a half of axion mass. As a result, a primary gravitational wave (GW) is followed by a secondary GW with a characteristic time delay and signal duration determined by the properties of axions, namely, their mass and coupling constant to the parity-violating sector of gravity.

In our previous paper, we obtained constraints on the effective coupling constant, which takes into account the fraction of axions in DM, using a method optimized for axion-induced GW signals~\cite{axion_tsutsui}. The constraint is $\sim 1\times 10^7\,\mathrm{km}$ for the axion mass range $[1.7 \times 10^{-13},\, 8.5 \times 10^{-12}] \,\mathrm{eV}$, which is about ten times stronger than that by Gravity Probe B~\cite{GravityProbeB, GravityProbeB_satellite} if the axions constitute the dominant DM component. However, the constraint was based on the unrealistic assumption that the DM density in the MW halo is homogeneous. In reality, the DM distribution is dense at the center of the MW halo and sparse at the edge. In this paper, we consider a realistic DM profile called the core NFW profile~\cite{coreNFW_profile1, coreNFW_profile2, coreNFW_profile3} and modify the properties of the axion-induced GW signals, resulting in a realistic constraint on the gravitational coupling of the axions.

The organization of this paper is as follows. In \secref{sec:review_of_the_previous_study}, we review the previous study~\cite{axion_tsutsui}. Next, a more realistic picture of the MW halo is explained in \secref{sec:pictures_in_the_milky_way_halo}. We then consider the modifications of the axion signals resulting from the difference in DM profiles in \secref{sec:properties_of_secondary_gravitational_waves}. In \secref{sec:conversion_of_the_previous_constraint}, we discuss the conversion of the previous constraint, taking the modifications into account~\cite{axion_tsutsui}. Finally, we conclude this study in \secref{sec:conclusion}. In this paper, we use the natural units $\hbar = c = 1$.

\section{Review of the previous study} \label{sec:review_of_the_previous_study}
In this section, we briefly review the previous search conducted in~\cite{axion_tsutsui}, which relied on the time delay and the amplification of a GW at a resonant frequency due to axion decay~\cite{soda-yoshida, soda-urakawa}.

\subsection{The properties of the secondary gravitational waves} \label{sec:the_properties_of_the_secondary_gravitational_waves}
The Lagrangians are
\eq{
	S =& S_{\mathrm{EH}} + S_{a} + S_{\mathrm{CS}}, \label{EQ:total_action} \\
	S_{\mathrm{EH}} =& \frac{1}{16 \pi G} \int d^4 x \sqrt{-g} R, \\
	S_a =& - \int d^4 x \sqrt{-g} \left( \frac{1}{2} \nabla_\mu a \nabla^\mu a + \frac{m^2_a}{2} a^2 \right), \\
	S_{\mathrm{CS}} =& \frac{\ell^2}{16 \sqrt{2 \pi G}} \int d^4 x \sqrt{-g}  a {^\ast\!} R R,
}
where $g$ is the determinant of the metric $g_{\mu\nu}$, $\ell$ is the axion coupling constant in the parity-violating sector of gravity, $m_a$ is the axion mass, $a$ is the axion field, and $^*RR$ is the Pontryagin density~\cite{review_CSgravity}.

By solving the equation of motion of the axion field in a homogeneous background, we obtain, with the complex amplitude $a_0$,
\eq{
	a(t) = \frac{a_0}{2} \e^{-\iu m_a t} + \mathrm{c.c.}\;.
}
The solution should be valid at the scale smaller than the de Broglie wavelength;
\eq{
	L_c =& \frac{2\pi}{m_a \varDelta v} \nonumber \\
		=& 4.0\times 10^{-8}\,\mathrm{pc} \typicalparam{1 \times 10^{-12}\,\mathrm{eV}}{m_a} \typicalparam{\varDelta v}{1\times 10^{-3}} \;, \label{eq:Lc}
}
where $\varDelta v$ is the DM velocity dispersion~\cite{MW_dark_matter_velocity_dispersion}.
We note that the factor of $2\pi$ is different from the definition in~\cite{soda-urakawa}.
Therefore, the axions form coherent clouds with the size of $L_c$. The number of axion patches along the line of sight is
\eq{
	N \coloneqq& R / L_c \nonumber \\
		=& 2.5\times 10^{12} \typicalparam{R}{100\,\mathrm{kpc}} \typicalparam{\varDelta v}{1\times 10^{-3}} \typicalparam{m_a}{1 \times 10^{-12}\,\mathrm{eV}}\;,
}
where $R$ is the radius of the MW halo.

The complex amplitude $a_0$ is related to the axion energy density as $\rho_a = m_a^2 \left| a_0 \right|^2 /2$, that is, $\left| a_0 \right| = \sqrt{2 \rho_a} / m_a$. As the axions might not be a dominant component in DM, we use the effective coupling constant 
\eq{
	\ell_\mathrm{eff} \coloneqq \ell f_\mathrm{DM}^{1/4} \;, \label{eq:effective_coupling}
}
where the fraction of the amount of axions to the total amount of the DM is $f_\mathrm{DM} \coloneqq \rho_a / \rho_\mathrm{DM}$. With the definition, we do not need to discriminate between the density of the axions and the DM, because the factor in the $S_\mathrm{CS}$ is $\ell^2 a \propto \ell^2 a_0 \propto \ell^2 \sqrt{\rho_a} = \ell_\mathrm{eff}^2 \sqrt{\rho_\mathrm{DM}}$. 

Since we consider axion DM in the Milky Way halo or at a much smaller scale than the cosmic scale, we neglect the cosmic expansion. That is, the metric is written as $g_{\mu\nu} = \eta_{\mu\nu} + h_{\mu\nu}$ where $\eta_{\mu\nu}$ is the Minkowski metric and $h_{\mu\nu}$ is the metric perturbation. By solving the equation of motion for $h_{\mu\nu}$, we find the solutions of the forward and the backward waves at the resonant frequency
\eq{
	f_\mathrm{res} \coloneqq& \frac{1}{2\pi} \frac{m_a}{2} \nonumber \\
		=& 1.2\times 10^2\,\mathrm{Hz} \typicalparam{m_a}{1 \times 10^{-12}\,\mathrm{eV}} \;, \label{eq:f_res}
}
with the width of the resonance
\eq{
	\varDelta f_\mathrm{res} \coloneqq& 2 f_\mathrm{res} \varDelta v \nonumber \\
		=& 0.24\,\mathrm{Hz} \typicalparam{m_a}{1 \times 10^{-12}\,\mathrm{eV}} \typicalparam{\varDelta v}{1\times 10^{-3}} \;. \label{eq:Delta_f_res}
}
As the width is narrow, the forward and backward waves can be considered nearly monochromatic. While the backward wave violates parity symmetry and is of interest, it is neglected due to its much smaller flux compared to the forward wave. The observed forward waveforms of the right- and left-hand modes are $h_{\mu\nu}^{R/L} = e_{\mu\nu}^{R/L} h_F^{R/L} + \mathrm{c.c.}$ with
\eq{
	h_F^{R/L}(t, \vec{x}) \coloneqq& h_{F, 0}^{R/L} F_\mathrm{total} \e^{\iu (\vec{k} \cdot \vec{x} - \omega t)} \;, \\
	F_\mathrm{total} \coloneqq& \prod_{i = 1}^N \left( 1 + \delta_\mathrm{patch} \right) \e^{\iu \psi_\mathrm{patch}}\;, \label{eq:Ftotal_naive}
}
where $e_{\mu\nu}^{R/L}$ is the polarization tensor, $h_{F, 0}^{R/L}$ is the complex amplitude including an initial phase, and $\delta_\mathrm{patch}$ and $\psi_\mathrm{patch}$ are the amplification factor and phase shift in a single patch. Note that the quantities with the subscript ``patch'' are those contributed by a single axion patch. The index $i$ is used to distinguish between different axion patches. However, in the previous paper~\cite{axion_tsutsui}, we did not consider this dependence due to the assumption of a homogeneous density profile, under which the amplification factor and phase shift are
\eq{
	\delta_\mathrm{patch} \coloneqq& \frac{\pi^3 G \varsigma}{2} m_a^2 \ell_\mathrm{eff}^4 \frac{\rho_\mathrm{DM}}{\varDelta v^2} \;, \label{eq:delta_patch} \\
	\psi_\mathrm{patch} \coloneqq& \frac{2\pi^4 G}{3} m_a^2 \ell_\mathrm{eff}^4 \frac{\rho_\mathrm{DM}}{\varDelta v^2} \frac{f - f_\mathrm{res}}{\varDelta f_\mathrm{res}} \;. \label{eq:psi_patch}
}
The $\varsigma$ is a parameter to average the frequency dependence of the signal amplitude in the frequency range between $f_\mathrm{res} \pm \varDelta f_\mathrm{res}$, i.e. $\varsigma \coloneqq \frac{1}{2} \int_{-1}^{+1} \sinc^2(2\pi x) \diff x = 0.24$ as estimated in the previous paper~\cite{axion_tsutsui}.

Due to the accumulation of the phase shift $\psi_\mathrm{patch}$ during propagation, the group velocity of the forward wave is delayed~\cite{axion_tsutsui, soda-urakawa}.
That is, a forward wave is observed as secondary GWs.
The time delay from a primary GW is
\eq{
	\varDelta t_\mathrm{delay} \coloneqq& N \varDelta t_\mathrm{delay}^\mathrm{patch} \;, \nonumber \\
	\varDelta t_\mathrm{delay}^\mathrm{patch} \coloneqq& \frac{2\pi^4 G}{3} m_a \ell_\mathrm{eff}^4 \frac{\rho_\mathrm{DM}}{\varDelta v^3} \;. \label{eq:delay}
}
The signal duration of the secondary GW is about
\eq{
	\varDelta t_\mathrm{duration} &\coloneqq 1 / \varDelta f_\mathrm{res} \;, \nonumber \\
	&= 4.1\,{\mathrm s} \typicalparam{1 \times 10^{-12}\,\mathrm{eV}}{m_a} \typicalparam{1\times 10^{-3}}{\varDelta v} \;. \label{eq:duration}
}

Since the secondary GWs have the frequency width $\varDelta f_\mathrm{res}$, the phase difference between the boundary modes $f_\mathrm{res} \pm \varDelta f_\mathrm{res}$ cannot be neglected \footnote{
	Although the derivation of $\varDelta \psi_\mathrm{patch}$ is different from that in~\cite{axion_tsutsui}, the result is consistent because of $\varDelta \psi_\mathrm{patch} = 2\pi (2 \varDelta f_\mathrm{res}) \varDelta t_\mathrm{delay}^\mathrm{patch}$.
} and is given by
\eq{
	\varDelta\psi \coloneqq& \sum_{i = 1}^N \left[ \left. \psi_\mathrm{patch} \right|_{f = f_\mathrm{res} + \varDelta f_\mathrm{res}} - \left. \psi_\mathrm{patch} \right|_{f = f_\mathrm{res} - \varDelta f_\mathrm{res}} \right] \nonumber \\
		=& N \varDelta \psi_\mathrm{patch} \;, \\
	\varDelta\psi_\mathrm{patch} \coloneqq& \frac{4\pi^4 G}{3} m_a^2 \ell_\mathrm{eff}^4 \frac{\rho_\mathrm{DM}}{\varDelta v^2} \;. \label{eq:delta_psi_patch}
}
By defining the critical number of patches $N_c$ with
\eq{
	\left. \varDelta \psi \right|_{N = N_c} = \pi \;, \label{eq:Nc}
}
the amplification is coherent for $N \leq N_c$ and incoherent for $N > N_c$. The total amplification factor $F_\mathrm{total}$ in Eq.~\eqref{eq:Ftotal_naive} is given by $F_\mathrm{total}^\mathrm{uni} = [(1 + \delta_\mathrm{patch}) \e^{\psi_\mathrm{patch}}]^N$. Because of $\delta_\mathrm{patch} \ll 1$, the coherent amplification converges to $\exp(N \delta_\mathrm{patch})$. The incoherent amplification is proportional to $\sqrt{N / N_c} = \sqrt{\varDelta \psi / \pi}$ because it is similar to the Brownian motion~\cite{axion_tsutsui}. In summary, the amplification can be written as
\eq{
	F_\mathrm{total}^\mathrm{uni} \simeq \left\{
		\begin{array}{ll}
		\displaystyle
		\e^{N \delta_\mathrm{patch}} \qquad \qquad {\rm for}\; N \leq N_{\rm c} \; (\mathrm{or\ \varDelta \psi \leq \pi}) \\
		\displaystyle
		\e^{N_{\rm c} \delta_\mathrm{patch}} \sqrt{\frac{\varDelta \psi}{\pi}} \quad {\rm for}\; N > N_{\rm c} \; (\mathrm{or\ \varDelta \psi > \pi})
	\end{array} \right. \;. \label{eq:Ftotal}
}
For the critical case $N = N_c$, the critical coupling is
\eq{
	\ell_c^\mathrm{eff} =& 8.0\times 10^6\,\mathrm{km} \typicalparam{\varDelta v}{1\times 10^{-3}}^{1/4} \typicalparam{100\,\mathrm{kpc}}{R}^{1/4} \nonumber \\
		&\typicalparam{1 \times 10^{-12}\,\mathrm{eV}}{m_a}^{3/4} \typicalparam{0.3\,\mathrm{GeV}/\mathrm{cm}^3}{\rho_\mathrm{DM}}^{1/4} \;.
}
When a primary GW is from a compact binary coalescence (CBC), the Fourier amplitudes of the axion signal $\tilde{h}_\mathrm{axion}$ is related to the Fourier waveform of a CBC, $\tilde{h}_\mathrm{CBC}$, as
\eq{
	\left| \tilde{h}_\mathrm{axion}(f_\mathrm{res}) \right| = \left( F_\mathrm{total} - 1 \right) \left| \tilde{h}_\mathrm{CBC}(f_\mathrm{res}) \right| \;. \label{eq:axion_signal_amplitude}
}
For $\tilde{h}_\mathrm{CBC}$ of binary black holes, we use the IMRPhenomD waveform~\cite{PNexample_3.5PN_1, PNexample_3.5PN_2}, which is an aligned spinning inspiral-merger-ringdown waveform, setting the high frequency cutoff to the peak frequency at which the amplitude of the waveform is maximized.
While for binary neutron stars, the waveform is not accurate enough at high frequencies due to tidal deformation, so the high-frequency cutoff is set to the frequency of the innermost stable circular orbit for a Schwarzschild black hole.

\subsection{Search method optimized for the axion signals} \label{sec:search_method_optimized_for_the_axion_signals}
By considering a set of $(m_a, \ell_\mathrm{eff})$, we can estimate, from Eqs.~\eqref{eq:delay} and~\eqref{eq:duration}, where the axion signals are expected to exist in the data obtained after the reported CBC mergers so far. Subsequently, by taking the corresponding data chunk $d$, whose duration is equal to $\varDelta t_\mathrm{duration}$, we can calculate the observed axion SNR,
\eq{
	\chi^2_\mathrm{obs} \coloneqq \sum_{I\in\mathrm{IFO}} \left| \frac{\tilde{d}_I(f_\mathrm{res})}{\sqrt{S_{n,I}(f_\mathrm{res})}} \sqrt{\varDelta f_\mathrm{res}} \right|^2 \;,
}
where $S_{n,I}(f)$ is the noise power spectral density of the $I$th detector.

The observed axion SNR obeys the noncentral $\chi^2$-distribution $p_\chi(x | \lambda)$ with degree of freedom equal to twice the number of GW detectors and the noncentral parameter $\lambda$, which is given by
\eq{
	\lambda \coloneqq \sum_{I\in\mathrm{IFO}} \left| \frac{\tilde{h}_\mathrm{axion}(f_\mathrm{res})}{\sqrt{S_{n,I}(f_\mathrm{res})}} \sqrt{\varDelta f_\mathrm{res}} \right|^2 \;. \label{eq:lambda}
}
This quantity is equivalent to the axion SNR in the absence of noise, so that $\lambda$ represents the expected axion SNR. Consequently, we can compute the $p$-value for a GW event as
\eq{
	p(m_a, \ell_\mathrm{eff}) \coloneqq \int_0^{\chi^2_\mathrm{obs}} p_\chi(x | \lambda) \, \diff x \;. \label{eq:p_value}
}
For instance, if the observed axion SNR is significantly smaller than the expected value, the corresponding $p$-value would be very small. In our previous study~\cite{axion_tsutsui}, we adopted a $p$-value threshold of $0.5\%$, and rejected parameter sets whose $p$-values were lower than this threshold.

After calculating the $p$-values for several GW events, it is necessary to combine the results. We assume that it is impossible to not observe the secondary GWs after all GW events, even if there are axion signals present. To express this assumption, we use the Logical OR to combine the constraint result for each primary GW event.

In the previous study~\cite{axion_tsutsui}, we utilized the observation data after GW170814, GW170817, GW190728\_064510, GW200202\_154313, and GW200316\_215756. We chose these events because they were detected by three GW detectors (i.e., two LIGO~\cite{LIGO1, LIGO2} and Virgo~\cite{Virgo}), and the data obtained after the detections were long and had high duty cycles. The searched mass range is $[1.7 \times 10^{-13},\, 8.5 \times 10^{-12}] \,\mathrm{eV}$, which corresponds to the LIGO sensitive band $[20,\,1024]\,\mathrm{Hz}$. The mass range is binned like $\varDelta \log m_a = \varDelta v$ so that the frequency bin size determined by Eq.~\eqref{eq:f_res} is equal to $\varDelta f_\mathrm{res}$. The effective coupling range that was searched for is determined so that the time delay in Eq.~\eqref{eq:delay} does not exceed the length of the searched data (see~\cite{axion_tsutsui} for the detail data range). The binning of the effective coupling is given by the solution of $\varDelta t_\mathrm{delay} = \alpha \varDelta t_\mathrm{duration}$ for the equal interval of  the shift ratio $\alpha$:
\eq{
	\ell_\mathrm{eff} =& 7.9\times 10^6\,\mathrm{km} \typicalparam{100\,\mathrm{kpc}}{R}^{1/4} \typicalparam{\varDelta v}{1\times 10^{-3}}^{1/4} \nonumber \\
		&\typicalparam{0.3\,\mathrm{GeV}/\mathrm{cm}^3}{\rho_\mathrm{DM}}^{1/4} \typicalparam{1 \times 10^{-12}\,\mathrm{eV}}{m_a}^{3/4} \typicalparam{\alpha}{1}^{1/4} \;. \label{eq:alpha_ell_relation}
}
The interval of the shift ratio is $\varDelta \alpha = 0.002$ for $\alpha \in (0.002, 0.2]$ and $\varDelta\alpha = 0.2$ for $\alpha \in (0.2, 4\time10^4]$, determined from the axion SNR loss~\cite{axion_tsutsui}.

\section{Pictures in the Milky Way halo} \label{sec:pictures_in_the_milky_way_halo}
In the previous study~\cite{axion_tsutsui} reviewed in \secref{sec:review_of_the_previous_study}, we assume a uniform DM profile and constant DM velocity dispersion in the MW halo. However, this assumption is unrealistic, and although we used typical values to estimate the properties of the axion signals, a more accurate representation of the MW halo is necessary. Therefore, in this section, we present a more realistic picture of the MW halo.

As DM particles are massive, they tend to be more densely distributed towards the center and sparsely distributed towards the edge of the MW halo. In this paper, we consider a more realistic DM profile known as the core NFW profile~\cite{coreNFW_profile1, coreNFW_profile2, coreNFW_profile3}:
\eq{
	\rho_\mathrm{cNFW}(r) \coloneqq& \frac{1}{4\pi r^2} \frac{\diff M_\mathrm{cNFW}(r)}{\diff r} \;, \\
	M_\mathrm{cNFW}(r) \coloneqq& M_\mathrm{NFW}(r) \tanh^n\left( \frac{r}{r_c} \right) \;, \\
	M_\mathrm{NFW}(r) \coloneqq& 4\pi \rho_s r_s^3 \left[ \log\left( 1 + \frac{r}{r_s} \right) + \frac{1}{1 + r / r_s} - 1 \right] \;,
}
where
\eq{
	n = \tanh\left( \kappa \frac{t_\mathrm{SF}}{t_\mathrm{dyn}} \right) \;.
}
The function $M_\mathrm{NFW}$ is the inner mass for the NFW profile~\cite{NFW_profile, NFW_profile_parameters}, $r_s$ is the scale radius, $r_c$ is the core size, $\kappa$ is the fitting parameter, $t_\mathrm{SF}$ is the total star formation time, and the circular orbit time at the $r_s$ is
\eq{
	t_\mathrm{dyn} = 2\pi \sqrt{\frac{r_s^3}{GM_\mathrm{NFW}(r_s)}} \;.
}
We choose those parameters as $r_s = 8.1\,\mathrm{kpc}$, $\rho_s = 0.51\,\mathrm{GeV}/\mathrm{cm}^3 \frac{R_\odot}{r_s} \left( 1 + \frac{R_\odot}{r_s} \right)^2$, $R_\odot = 8.0\,\mathrm{kpc}$, $t_\mathrm{SF} = 1.4\times 10^{10}\,{\rm year}$, $r_c = 0.175\,\mathrm{kpc}$, and $\kappa = 0.04$~\cite{NFW_profile_parameters, coreNFW_profile2, coreNFW_profile3}. 

\begin{figure}
	\centering
	\includegraphics[width=0.95\linewidth]{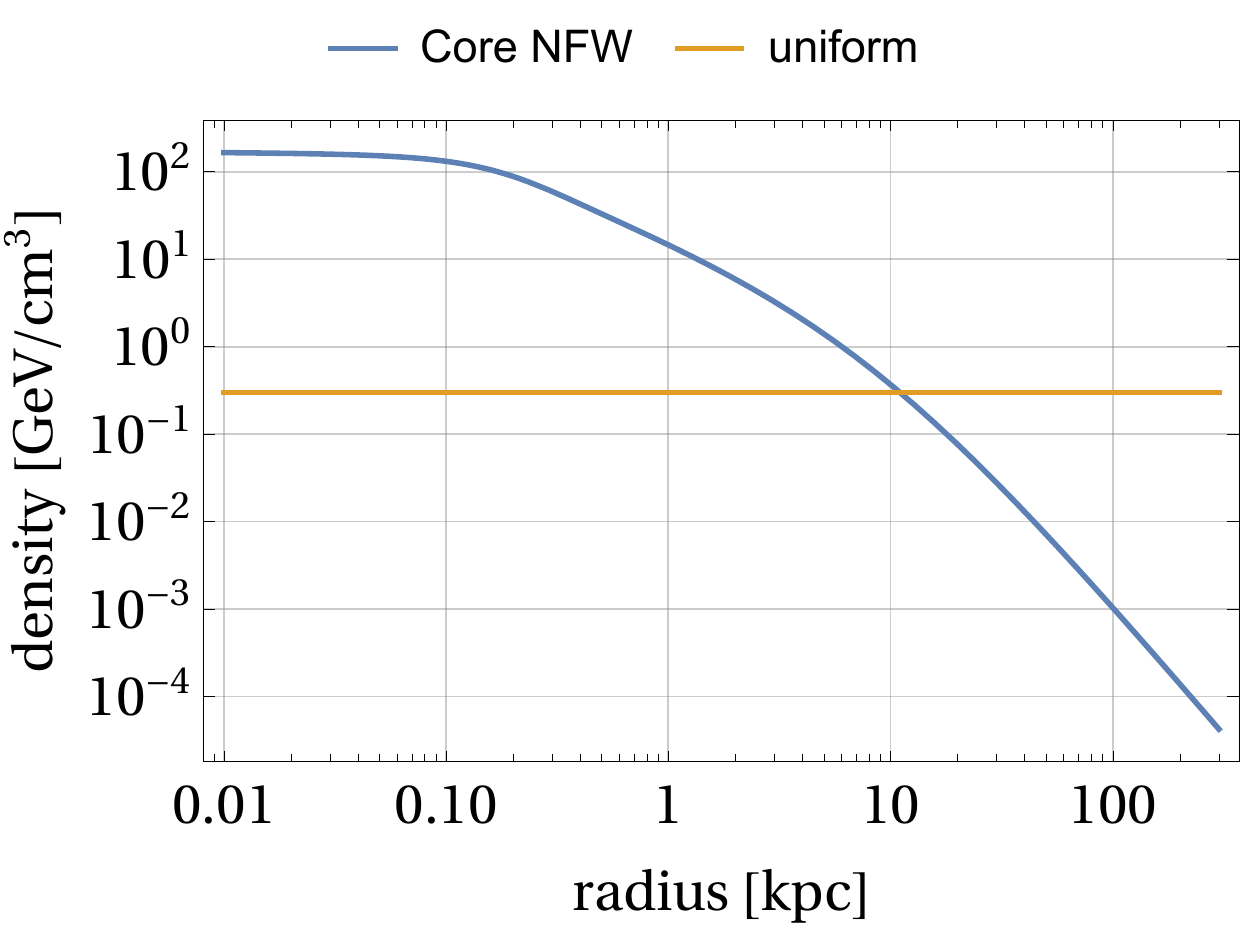}
	\caption{The DM density profiles. The blue line represents the core NFW profile, $\rho_\mathrm{cNFW}(r)$, while the orange one is the uniform profile assumed in the previous study~\cite{axion_tsutsui}, or \secref{sec:review_of_the_previous_study}, for comparison.}
	\label{fig:density_profile}
\end{figure}

Figure~\ref{fig:density_profile} shows the density profiles for the core NFW profile and the uniform profile for comparison. The core NFW profile has a nearly constant core. This is the reason why we use the core NFW profile instead of the NFW profile which has the cusp core not favored by observed rotation curves~\cite{cusp-core1, cusp-core2, cusp-core3, cusp-core4, cusp-core5, cusp-core6, cusp-core7}. In addition, we cannot neglect the separation between the Earth and the Galactic center as was done in the previous study~\cite{axion_tsutsui}, as the density varies significantly between those locations.

Although the inner mass of the core NFW profile diverges for $r \to \infty$, it is unrealistic. Thus, an upper cutoff of $R_\mathrm{max}$ must be applied. Since the mass of the MW halo is about $1 \times 10^{12}\,M_\odot$ from observations~\cite{MWhalo_mass_observation1, MWhalo_mass_observation2, MWhalo_mass_observation3, MWhalo_mass_observation4}, we set $R_\mathrm{max} = 300\,\mathrm{kpc}$ because of $M_\mathrm{cNFW}(300\,\mathrm{kpc}) = 1.1\times 10^{12}\,M_\odot$.

The inner mass of the core NFW profile approaches zero as $r$ approaches zero, but this leads to divergence in the later calculation. Furthermore, a realistic galaxy is likely to have a supermassive black hole at its center. Hence, we set a lower cutoff of $R_\mathrm{min} = 0.1,\mathrm{kpc}$. This lower cutoff corresponds to neglecting a small region of the sky, approximately $10^{-3}\%$ of the celestial sphere, seen from the Earth. Considering the anisotropic properties of the axion signals later, the neglected region is sufficiently small, and the difference in the inner mass is also less than $10^{-3}\%$. Therefore, setting the lower cutoff has a negligible effect.

\begin{figure}
	\centering
	\includegraphics[width=0.95\linewidth]{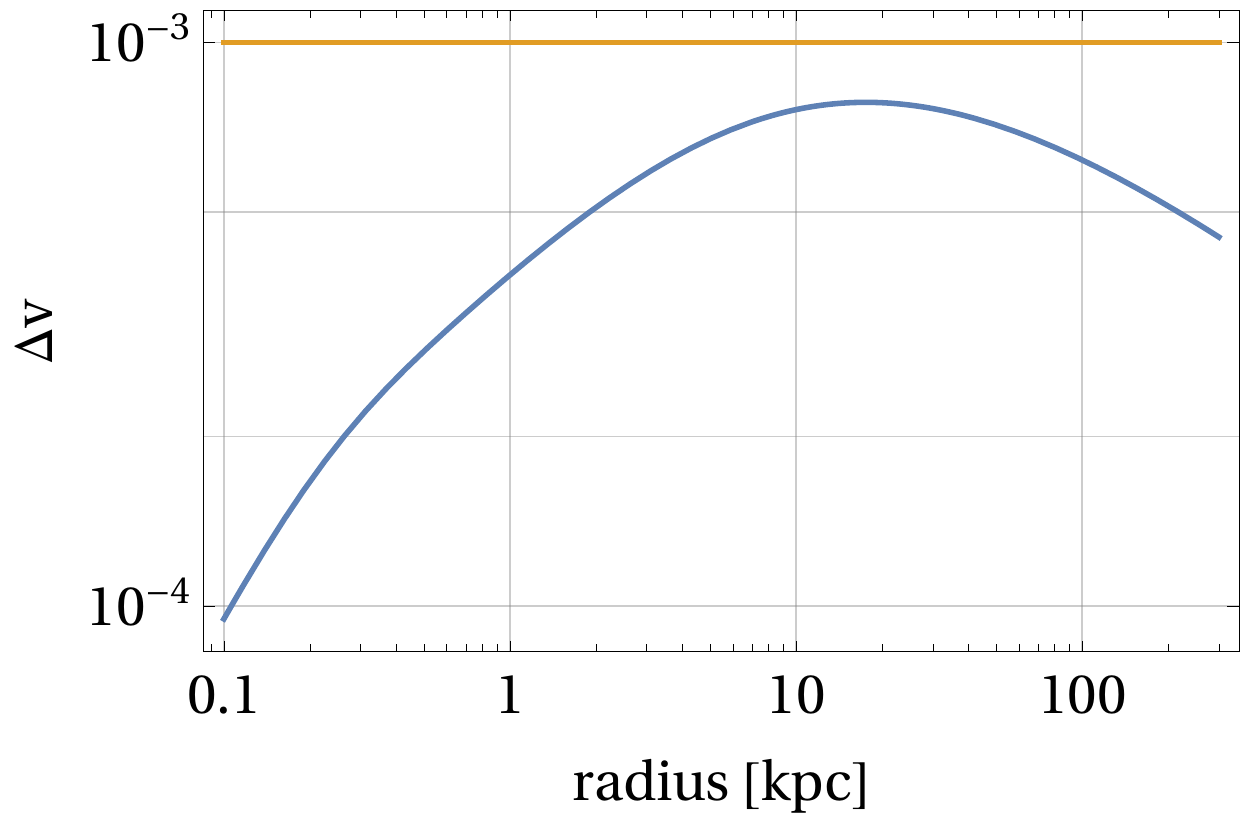}
	\caption{The velocity dispersion as a function of the radius of the MW halo. The blue line represents the core NFW profile, $\varDelta v^\mathrm{cNFW}(r)$, while the orange one is the constant velocity assumed in the previous study~\cite{axion_tsutsui}, or \secref{sec:review_of_the_previous_study}, for comparison.}
	\label{fig:velocity_dispersion}
\end{figure}

The velocity dispersion of DM is estimated based on its kinetic energy, which is related to its potential energy through the Virial theorem. Therefore, the velocity dispersion for the core NFW profile can be obtained as
\eq{
	\varDelta v^\mathrm{cNFW}(r) = \sqrt{\frac{6}{5} \frac{G M_\mathrm{cNFW}(r)}{r}} \;, \label{eq:velocity_dispersion_NFW}
}
which is shown in \figref{fig:velocity_dispersion}.
We note that the derivation using the Virial theorem has not yet been verified to be correct in observations and simulations due to the effects of star formation and supernovae, and the lower resolution of simulations, among other factors~\cite{review_axion}. In \secref{sec:conversion_of_the_previous_constraint}, we will discuss the robustness of our constraints.

In \figref{fig:velocity_dispersion}, there exists the maximized radius of $\varDelta v^\mathrm{cNFW}(r)$. However, it is not easy to derive the analytical form of the maximized radius because of the complexity of the inner mass $M_\mathrm{cNFW}(r)$. However, as the maximized radius is only required for an approximate estimation later, we take the maximized radius as $R_\mathrm{ext} = 20\,\mathrm{kpc}$ based on \figref{fig:velocity_dispersion}.

\section{Properties of secondary gravitational waves} \label{sec:properties_of_secondary_gravitational_waves}
In the previous section, we discussed how the DM density profile and velocity dispersion of the MW halo depend on its radius. Consequently, we modify the properties of the secondary GWs in the uniform density profile that reviewed in \secref{sec:review_of_the_previous_study}. Actually, the modifications of the properties of the axion signals have been studied in~\cite{axion_constraint_NFW}. The authors assume the NFW profile but not the core NFW profile. In addition, their derivations are only for the coherent case, which is not always true, particularly in the regime of strong coupling. Thus, we derive the properties of secondary GWs exhaustively, including a new result on incoherent amplification.

\subsection{Number of axion patches}
We can obtain the physical quantities for the core NFW profile by replacing $\rho_\mathrm{DM}$ and $\varDelta v$ with $\rho_\mathrm{cNFW}(r)$ and $\varDelta v^\mathrm{cNFW}(r)$, respectively, since the equations of motion are not affected by the density profile and velocity dispersion. Using $L_c$ in Eq.~\eqref{eq:Lc}, we obtain the coherent length:
\eq{
	L_c^\mathrm{cNFW}(r) \coloneqq \frac{2\pi}{m_a \varDelta v^\mathrm{cNFW}(r)} \;.
}

Since the coherent length depends on the radius for the core NFW profile, the number of patches along the path of GW propagation is also modified. Additionally, we consider the separation between the Earth and the Galactic center, which causes the modified number of patches to depend on the direction of the propagating GW. The direction is parameterized with the angle $\phi \coloneqq \arccos(\vec{n}_\mathrm{GW} \cdot \vec{n}_\mathrm{Earth})$ where $\vec{n}_\mathrm{GW}$ and $\vec{n}_\mathrm{Earth}$ are the unit vectors from the Galactic center to the GW source and the Earth, respectively. We assume that the Earth is on the line of $\phi = 0$. By definition, the Galactic core is in the direction of $\phi = \pi$ (see \appref{sec:jacobian_from_the_radius_to_the_propagating_ratio_of_the_path}). Since the MW halo is spherically symmetric and the propagation effect observed on the Earth is axially symmetric around the line of $\phi = 0$ and $\phi = \pi$, the angle around the axis is not considered in this paper. Because of $L_c^\mathrm{cNFW} \ll R_\mathrm{max}$, the number of axion patches can be approximated by integration:
\eq{
	N^\mathrm{cNFW}(\phi) \simeq& \int_{r \in P(\phi)} \left| \frac{\diff r}{L_c^\mathrm{cNFW}(r)} \right| \nonumber \\
		=& \int_0^1 \left| \frac{\diff r}{\diff x}(x, \phi) \right| \frac{\diff x}{L_c^\mathrm{cNFW}(r(x, \phi))} \;, \label{eq:N_NFW}
}
\eq{
	&\frac{\diff r}{\diff x}(x, \phi) \nonumber \\
		=& R_\mathrm{max} \frac{-(1 - x) + x \left( \frac{R_\odot}{R_\mathrm{max}} \right)^2 + (1 - 2x) \frac{R_\odot}{R_\mathrm{max}} \cos\phi}{\sqrt{(1 - x)^2 + x^2 \left( \frac{R_\odot}{R_\mathrm{max}} \right)^2 + 2x(1 - x) \frac{R_\odot}{R_\mathrm{max}} \cos\phi}} \;, \label{eq:dr_dx}
}
where the integral is computed along the path $P(\phi)$ of the propagating GWs in the MW halo.
The integral variable in Eq.~\eqref{eq:N_NFW} is changed from the radius $r$ to the affine coordinate along the path $x \in [0, 1]$ with Eq.~\eqref{eq:dr_dx} derived in \appref{sec:jacobian_from_the_radius_to_the_propagating_ratio_of_the_path}.
The intersection of the propagation path and the edge of the MW halo is at $x = 0$, and the Earth is at $x = 1$.

\subsection{Amplification} \label{sec:amplification}
From $\delta_\mathrm{patch}$ in Eq.~\eqref{eq:delta_patch}, the amplification factor from one axion patch in the core NFW profile is
\eq{
	\delta_\mathrm{patch}^\mathrm{cNFW}(r) \coloneqq& \frac{\pi^3 G \varsigma}{2} m_a^2 \ell_\mathrm{eff}^4 \frac{\rho_\mathrm{cNFW}(r)}{\varDelta v^\mathrm{cNFW}(r)^2} \;, \label{eq:delta_patch_NFW}
}
and the resonance width is, from $\varDelta f_\mathrm{res}$ in Eq.~\eqref{eq:Delta_f_res},
\eq{
	\varDelta f_\mathrm{res}^\mathrm{cNFW}(r) \coloneqq 2 f_\mathrm{res} \varDelta v^\mathrm{cNFW}(r) \;.
 \label{eq:resonance-freq-width}
}
Although the averaging parameter $\varsigma$ depends on $r$ through $\varDelta f_\mathrm{res}^\mathrm{cNFW}$ (see the discussion below Eq.~\eqref{eq:psi_patch}), we assume that $\varsigma$ is constant and is equal to that for the uniform profile, i.e. $\varsigma = 0.24$~\cite{axion_tsutsui}, by taking $\varDelta v \approx \varDelta v^\mathrm{cNFW}(R_\mathrm{ext})$ (see \figref{fig:velocity_dispersion}). Furthermore, the $\varsigma$ affects the SNR of the axion signal, but our search is not very sensitive to the SNR (see Supplemental material in~\cite{axion_tsutsui}). Thus, the $\varsigma$ does not have to be treated accurately.

\begin{figure}
	\centering
	\includegraphics[width=0.95\linewidth]{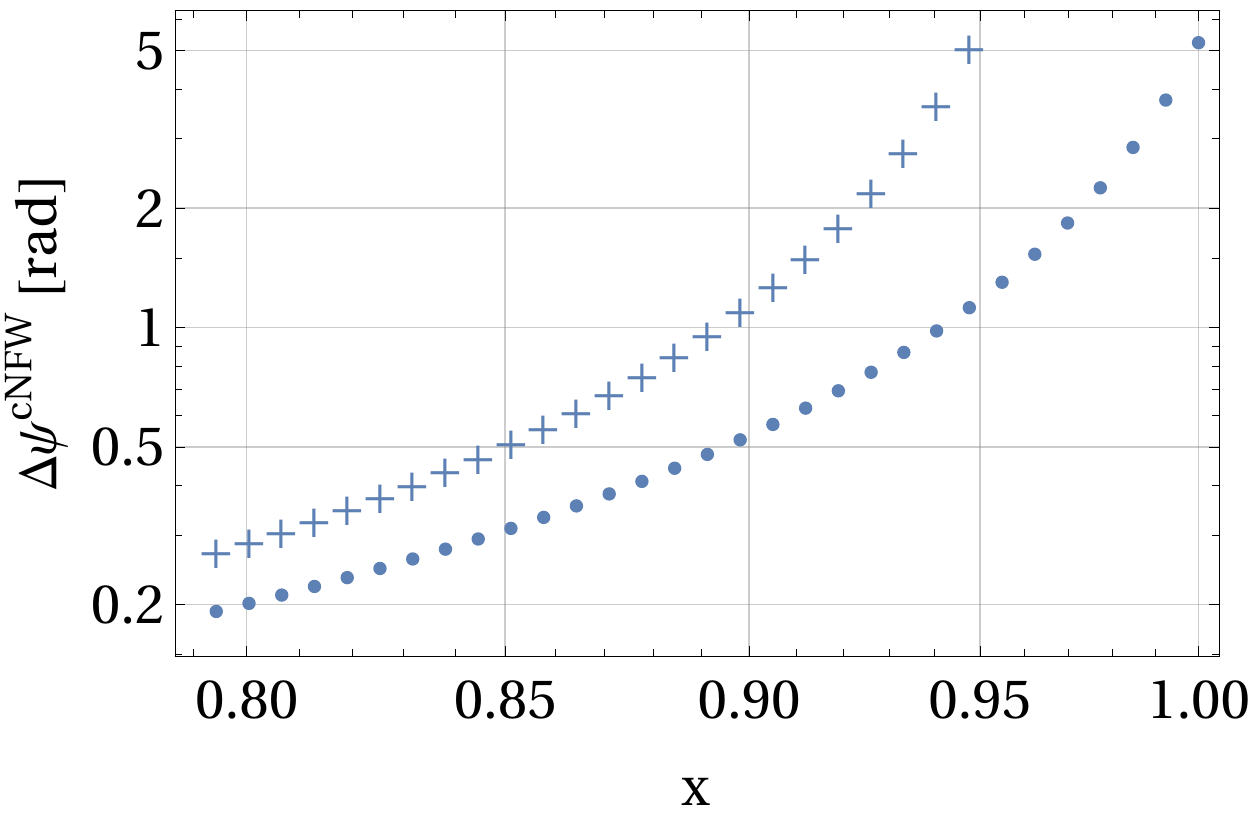}
	\caption{The accumulated phase difference, $\varDelta \psi^\mathrm{cNFW}(r(x), \phi)$, as a function of the affine coordinate along the path. The dot marker is for $\phi = 0$, and the plus marker is for $\phi = \pi - 1.01 \varDelta \phi_\mathrm{GC}$.  To plot those, the parameters are assumed to be $m_a = 1 \times 10^{-12}\,\mathrm{eV}$ and $\ell_\mathrm{eff} = 1\times 10^7\,\mathrm{km}$.}
	\label{fig:phase_shift}
\end{figure}

In order to calculate the amplification accumulated over a given propagation path, it is necessary to take into account the phase shift that accumulates at each axion patch, and to analyze the transition from coherent to incoherent amplification. From $\varDelta \psi_\mathrm{patch}$ in Eq.~\eqref{eq:delta_psi_patch}, the phase difference of the secondary GWs between the frequencies, $f_\mathrm{res} \pm \varDelta f_\mathrm{res}$ in a single axion patch is
\eq{
	\varDelta \psi_\mathrm{patch}^\mathrm{cNFW}(r) \coloneqq \frac{4 \pi^4 G}{3} m_a^2 \ell_\mathrm{eff}^4 \frac{\rho_\mathrm{cNFW}(r)}{\varDelta v^\mathrm{cNFW}(r)^2} \;. \label{eq:delta_psi_patch_NFW}
}
Then the accumulated phase shift of the GWs during propagation in the MW halo is
\eq{
	& \varDelta \psi^\mathrm{cNFW}(r, \phi) \coloneqq \int_{r' \in P'(r, \phi)} \varDelta \psi_\mathrm{patch}^\mathrm{cNFW}(r') \left| \frac{\diff r'}{L_c^\mathrm{cNFW}(r')} \right| \\
		=& \frac{2\pi^3 G}{3} m_a^3 \ell_\mathrm{eff}^4 \int_0^{x(r, \phi)} \frac{\rho_\mathrm{cNFW}(r'(x, \phi))}{\varDelta v^\mathrm{cNFW}(r'(x, \phi))} \left| \frac{\diff r'}{\diff x}(x, \phi) \right| \diff x
}
where $P'(r, \phi)$ is the path with a fixed $\phi$ of the propagating GWs from the edge of the MW halo at $R_\mathrm{max}$ to the point with the radius $r$ on the path.
Figure~\ref{fig:phase_shift} shows the accumulated phase shifts $\varDelta \psi^\mathrm{cNFW}(r(x), \phi)$ as a function of $x$ for $\phi = 0$ and $\phi = \pi - 1.01 \varDelta \phi_\mathrm{GC}$, where $\varDelta \phi_\mathrm{GC} \coloneqq \arctan \left( \frac{R_\mathrm{min}}{R_\odot} \right) \simeq 0.012\,{\rm rad}$.
For other angles of $\phi$, the accumulated phase differences are between the two cases.

From the condition, $\left. \varDelta \psi^\mathrm{cNFW} \right|_{N = N_c^\mathrm{cNFW}} = \pi$,
\eq{
	& N_c^\mathrm{cNFW}(\phi) \nonumber \\
		\coloneqq& \int_0^1 \left| \frac{\diff r}{\diff x}(x, \phi) \right| \frac{\Theta(\pi - \varDelta\psi^\mathrm{cNFW}(r(x, \phi), \phi))}{L_c^\mathrm{cNFW}(r(x, \phi))} \diff x
}
where $\Theta(\cdot)$ is the step function, which guarantees that the number of patches is counted up to $\varDelta\psi^\mathrm{cNFW} = \pi$.

So far, all the ingredients to construct the accumulated amplification have been prepared.
The observed coherent amplification factor at the Earth is, from $F_\mathrm{total}$ in Eq.~\eqref{eq:Ftotal_naive},
\eq{
	& F_\mathrm{total}^\mathrm{cNFW}(\phi) \coloneqq \prod_{i = 1}^{N^\mathrm{cNFW}(\phi)} \left[ 1 + \delta_\mathrm{patch}^\mathrm{cNFW}(r_i) \right] \nonumber \\
		\simeq& \exp\left[ \sum_{i = 1}^{N^\mathrm{cNFW}(\phi)} \delta_\mathrm{patch}^\mathrm{cNFW}(r_i) \right] \nonumber \\
		\simeq& \exp\left[ \int_{r \in P(\phi)} \left| \frac{\diff r}{L_c^\mathrm{cNFW}(r)} \right| \delta_\mathrm{patch}^\mathrm{cNFW}(r) \right] \nonumber \\
		=& \exp\left[ \frac{\pi^2 G \varsigma}{4} m_a^3 \ell_\mathrm{eff}^4 \int_0^1 \left| \frac{\diff r}{\diff x}(x, \phi) \right| \frac{\rho_\mathrm{cNFW}(r(x, \phi))}{\varDelta v^\mathrm{cNFW}(r(x, \phi))} \diff x \right] \;. \label{eq:Ftotal_NFW_coherent}
}
This expression for $F_\mathrm{total}$ is only applicable for coherent amplification when $N^\mathrm{cNFW}(\phi) \leq N_c^\mathrm{cNFW}(\phi)$, meaning $\varDelta \psi^\mathrm{cNFW}(r(x = 1, \phi), \phi) \leq \pi$.

Next, we consider incoherent amplification. Once the phase difference is accumulated to $\pi$, the amplification becomes incoherent like Brownian motion as discussed in \secref{sec:the_properties_of_the_secondary_gravitational_waves}. The amplification factor taken into account incoherent propagation is obtained from $F_\mathrm{total}^\mathrm{uni}$ in Eq.~\eqref{eq:Ftotal}:
\eq{
	F_\mathrm{total}^\mathrm{cNFW}(\phi) = F_\mathrm{total, c}^\mathrm{cNFW}(\phi) \sqrt{\frac{\varDelta \psi^\mathrm{cNFW}(r(x = 1, \phi), \phi)}{\pi}} \;, \label{eq:Ftotal_NFW_incoherent}
}
where, from the coherent amplification in Eq.~\eqref{eq:Ftotal_NFW_coherent},
\eq{
	& \log F_\mathrm{total, c}^\mathrm{cNFW}(\phi) \coloneqq \int_0^1 \left| \frac{\diff r}{\diff x}(x, \phi) \right| \frac{\delta_\mathrm{patch}^\mathrm{cNFW}(r(x, \phi))}{L_c^\mathrm{cNFW}(r(x, \phi))} \nonumber \\ 
 & \qquad \qquad \qquad \times \Theta(\pi - \varDelta\psi^\mathrm{cNFW}(r(x, \phi), \phi)) \diff x \;. \label{eq:logFc_NFW}
}
This expression is only valid for $N^\mathrm{cNFW}(\phi) > N_c^\mathrm{cNFW}(\phi)$, that is, $\varDelta \psi^\mathrm{cNFW}(r(x = 1, \phi), \phi) > \pi$.

\begin{figure}
	\centering
	\includegraphics[width=0.95\linewidth]{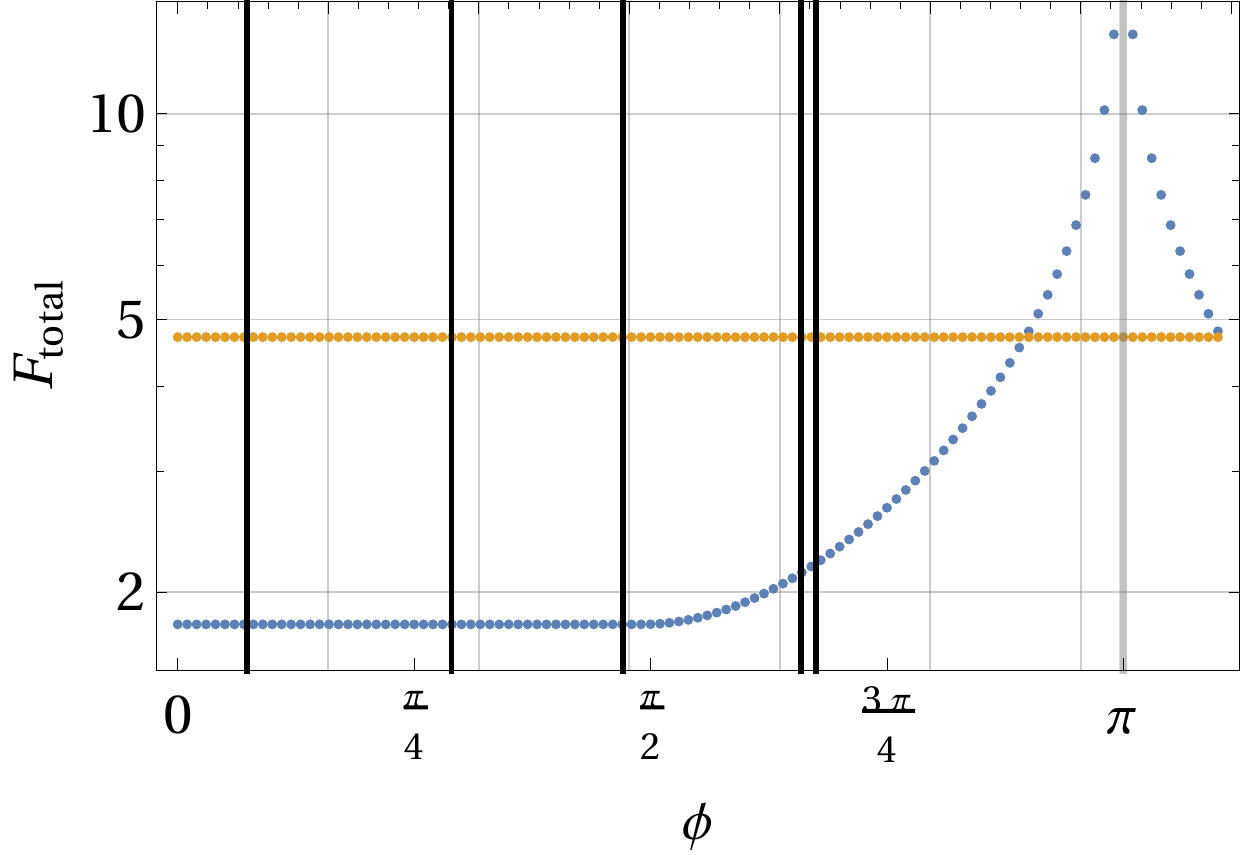}
	\caption{The amplification factor as a function of $\phi$.  The blue dots are $F_\mathrm{total}^\mathrm{cNFW}(\phi)$ in Eqs.~\eqref{eq:Ftotal_NFW_coherent} or~\eqref{eq:Ftotal_NFW_incoherent}.  The orange ones are $F_\mathrm{total}^\mathrm{uni}$ in Eq.~\eqref{eq:Ftotal} for comparison.  To plot those dots, we assume $m_a = 1 \times 10^{-12}\,\mathrm{eV}$ and $\ell_\mathrm{eff} = 1\times 10^7\,\mathrm{km}$.  The black vertical lines are the angles corresponding to the propagation directions of the analyzed primary GW events~\cite{axion_tsutsui}.  The gray region is inside the lower cutoff $R_\mathrm{min}$, that is, corresponds to the Galactic core.  Nothing is calculated in the gray region.}
	\label{fig:Ftotal_NFW}
\end{figure}

Figure~\ref{fig:Ftotal_NFW} shows the $\phi$-dependence of the amplification factor for $m_a = 1 \times 10^{-12}\,\mathrm{eV}$ and $\ell_\mathrm{eff} = 1\times 10^7\,\mathrm{km}$. Since GWs propagate in a denser region of the core NFW profile for $|\phi - \pi| \lesssim \pi / 2$, the amplification factor is also larger around the Galactic core. $F_\mathrm{total}^\mathrm{cNFW}$ exceeds $F_\mathrm{total}^\mathrm{uni}$ in the range of $|\phi - \pi| \lesssim 0.10 \pi$ for the parameter set of $m_a$ and $\ell_\mathrm{eff}$ in \figref{fig:Ftotal_NFW}. The range of $\phi$ for which $F_\mathrm{total}^\mathrm{cNFW} > F_\mathrm{total}^\mathrm{uni}$ is dependent on the chosen values of $m_a$ and $\ell_\mathrm{eff}$ through $F_\mathrm{total}^\mathrm{cNFW}(\phi)$, $ F_\mathrm{total}^\mathrm{uni}$, and $\varDelta\psi^\mathrm{cNFW}(\phi)$. Because of its complexity, it is not easy to derive the explicit dependence. However, as the $\phi$-range is determined mainly by the size of the Galactic core, its variation is about $\mathcal{O}(1\%)$ in the parameter region examined in this paper, for example, $\sim 2\%$ for $m_a = 1 \times 10^{-11},\mathrm{eV}$ and $\ell_\mathrm{eff} = 1\times 10^7,\mathrm{km}$, and $\sim 3\%$ for $m_a = 1 \times 10^{-13},\mathrm{eV}$ and $\ell_\mathrm{eff} = 1\times 10^8,\mathrm{km}$. Therefore, we can consider the $\phi$-range to be roughly universal in the examined parameter region. The propagation directions for which $F_\mathrm{total}^\mathrm{cNFW}$ is greater than $F_\mathrm{total}^\mathrm{uni}$ are $\sim 1\%$ of the celestial sphere. Assuming that GW sources are isotropically distributed in the Universe, only about $1\%$ of all GW events detected so far is expected to have $F_\mathrm{total}^\mathrm{cNFW} > F_\mathrm{total}^\mathrm{uni}$. For the reason, we choose events detected by at least three detectors with relatively large SNRs rather than expecting to have such a golden event.

\subsection{Time delay}
The time delay in a single axion patch at the radius $r$ is modified from Eq.~\eqref{eq:delay} as 
\eq{
	\varDelta t_\mathrm{delay}^\mathrm{patch, cNFW}(r) \coloneqq \frac{2\pi^4 G}{3} m_a \ell_\mathrm{eff}^4 \frac{\rho_\mathrm{cNFW}(r)}{\varDelta v^\mathrm{cNFW}(r)^3} \;. \label{eq:delay_patch_NFW}
}

The observed time delay at the Earth is the accumulation of tiny time delays from all the axion patches and is given by
\eq{
	& \varDelta t_\mathrm{delay}^\mathrm{cNFW}(\phi) \coloneqq \sum_{i = 1}^{N^\mathrm{cNFW}(\phi)} \varDelta t_\mathrm{delay}^\mathrm{patch, cNFW}(r_i) \nonumber \\
		\simeq& \int_{r \in D(\phi)} \varDelta t_\mathrm{delay}^\mathrm{patch, cNFW}(r) \left| \frac{\diff r}{L_c^\mathrm{cNFW}(r)} \right| \nonumber \\
		=& \frac{\pi^3 G}{3} m_a^2 \ell_\mathrm{eff}^4 \int_0^1 \left| \frac{\diff r}{\diff x}(x, \phi) \right| \frac{\rho_\mathrm{cNFW}(r(x, \phi))}{\varDelta v^\mathrm{cNFW}(r(x, \phi))^2} \diff x \label{eq:delay_NFW}
}
where $r_i$ is the radius at the $i$-th axion patch from the edge of the MW halo.

\begin{figure}
	\centering
	\includegraphics[width=0.95\linewidth]{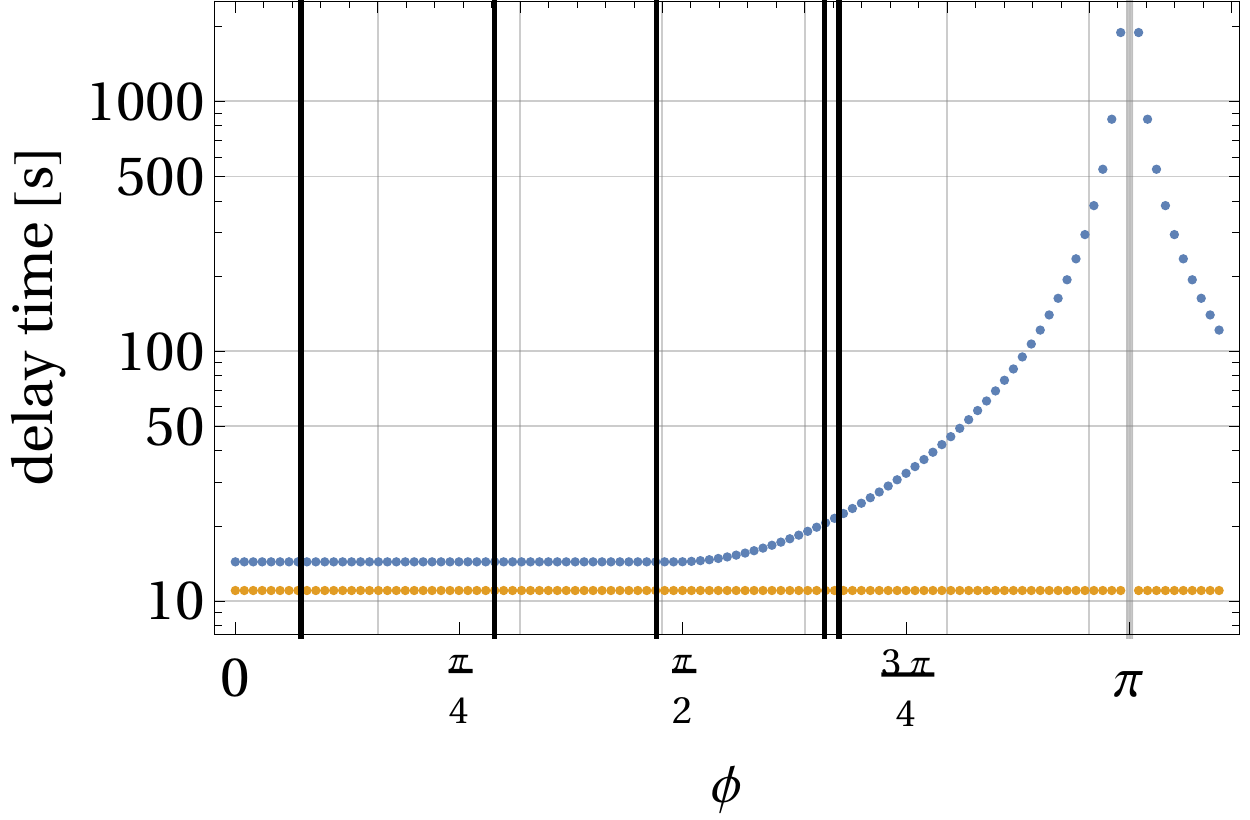}
	\caption{The time delay as a function of the $\phi$.  The blue dots are $\varDelta t_\mathrm{delay}^\mathrm{cNFW}(\phi)$ in Eq.~\eqref{eq:delay_NFW}. The orange ones are $\varDelta t_\mathrm{delay}$ in Eq.~\eqref{eq:delay} for comparison.  To plot those dots, we assume $m_a = 1 \times 10^{-12}\,\mathrm{eV}$ and $\ell_\mathrm{eff} = 1\times 10^7\,\mathrm{km}$. The black vertical lines are the angles corresponding to the propagation directions of the analyzed primary GW events~\cite{axion_tsutsui}. The gray region is inside the lower cutoff $R_\mathrm{min}$, that is, corresponds to the Galactic core.  Nothing is calculated in the gray region.}
	\label{fig:delay_NFW}
\end{figure}

Figure~\ref{fig:delay_NFW} illustrates the $\phi$-dependence of the time delay for $m_a = 1 \times 10^{-12},\mathrm{eV}$ and $\ell_\mathrm{eff} = 1\times 10^7,\mathrm{km}$. While the time delay in the core NFW profile is almost the same as that used in \secref{sec:the_properties_of_the_secondary_gravitational_waves} for $|\phi| \lesssim \pi / 2$, the difference becomes significant for $|\phi - \pi| \lesssim \pi / 2$. This is because the DM density of the core NFW profile around the Galactic center is much denser than that around the edge (see \figref{fig:density_profile}).

\subsection{Signal duration}
\begin{figure}
	\centering
	\includegraphics[width=0.95\linewidth]{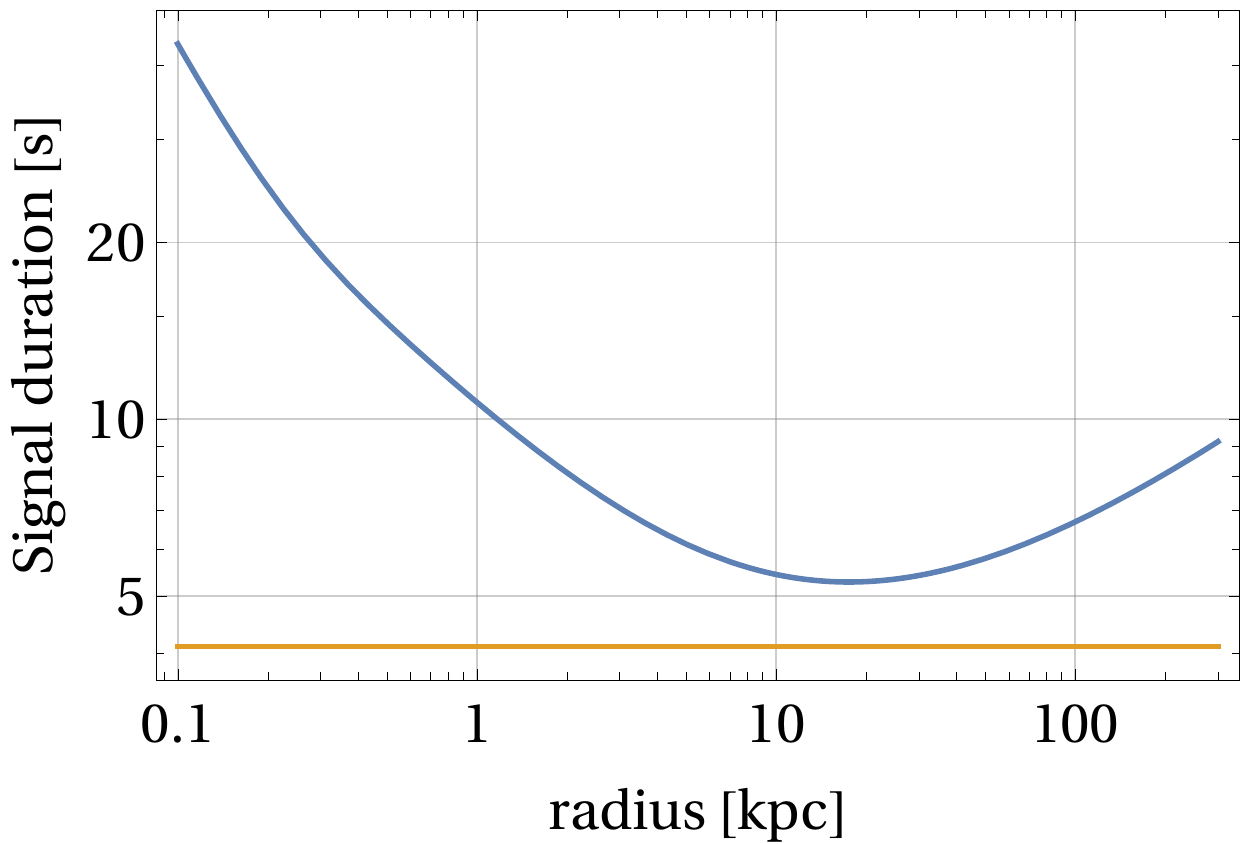}
	\caption{The signal duration when an axion signal is observed at a radius from the Galactic center. The blue line is $\varDelta t_\mathrm{duration}^\mathrm{cNFW}(r)$ in Eq.~\eqref{eq:duration_NFW}.  The orange one is $\varDelta t_\mathrm{duration}$ in Eq.~\eqref{eq:duration} for comparison.  To plot those lines, we assume $m_a = 1 \times 10^{-12}\,\mathrm{eV}$.}
	\label{fig:signal_duration_NFW}
\end{figure}

From $\varDelta t_\mathrm{duration}$ in Eq.~\eqref{eq:duration}, the duration of the axion signal for the core NFW profile is
\eq{
	\varDelta t_\mathrm{duration}^\mathrm{cNFW}(r) \coloneqq 1 / \varDelta f_\mathrm{res}^\mathrm{cNFW}(r) \;. \label{eq:duration_NFW}
}
Figure~\ref{fig:signal_duration_NFW} shows the signal duration for the core NFW profile when the axion signal is observed at the radius $r$ from the Galactic center.

Since we assume $\varDelta v \simeq \varDelta v^\mathrm{cNFW}(R_\mathrm{ext})$ in \secref{sec:amplification}, the observed signal duration does not depend on the angle $\phi$.
Even if the secondary GWs propagate over the Galactic core (i.e. $\phi = \pi$), the frequency width in Eq.~\eqref{eq:resonance-freq-width} is determined by the velocity dispersion at $r = R_\mathrm{ext} = 20\,\mathrm{kpc}$, where the velocity dispersion is maximized.

\section{Conversion of the previous constraint} 
\label{sec:conversion_of_the_previous_constraint}
In this section, based on the modified properties of secondary GWs in the case of the core NFW density profile derived above, we convert the previous constraint obtained for the uniform density profile into the one for the core NFW profile.

\begin{table}
	\centering
	\begin{tabular}{l|ccc} \hline\hline
		Event name       & $\phi$ & $\ell_\mathrm{eff}^\mathrm{cNFW} / 1\times 10^6\,\mathrm{km}$ & \shortstack{Relative Difference \\from Eq.~\eqref{eq:alpha_ell_relation}} \\ \hline
		GW170814         & $1.48$ & $7.79$                            & $-0.80\%$ \\
		GW170817         & $2.07$ & $7.13$                            & $-9.3\%$ \\
		GW190728\_064510 & $2.12$ & $7.00$                            & $-11\%$ \\
		GW200202\_154313 & $0.91$ & $7.79$                            & $-0.80\%$ \\
		GW200316\_215756 & $0.23$ & $7.79$                            & $-0.80\%$ \\ \hline
	\end{tabular}
	\caption{The relative differences of the effective coupling constants between the DM profiles, $(\ell_\mathrm{eff}^\mathrm{cNFW} - \ell_\mathrm{eff}^\mathrm{uni}) / \ell_\mathrm{eff}^\mathrm{uni}$, for the shift ratio $\alpha = 1$ and $m_a = 1 \times 10^{-12}\,\mathrm{eV}$ when the signal duration, $\varDelta t_\mathrm{duration}$, and the time delay modified in Eq.~\eqref{eq:delay_NFW} are used (the third column). As the mass dependence between $\ell_\mathrm{eff}^\mathrm{cNFW}$ and $\ell_\mathrm{eff}^\mathrm{uni}$ are common, the values of the relative difference are the same for other masses. In the first and second columns, we list the propagation directions~\cite{GWOSC} and the effective coupling constant estimated for the core NFW profile just for reference.}	\label{tab:relative_difference_bw_DMprofiles}
\end{table}

As the time delay and signal duration are affected by changes in the DM profile, the search region on the $m_a$--$\ell_\mathrm{eff}$ plane must be adjusted accordingly. We determine this region by solving $\varDelta t_\mathrm{delay}^\mathrm{cNFW} = \alpha \varDelta t_\mathrm{duration}^\mathrm{cNFW}$ with the same range of $\alpha$ as explained below Eq.~\eqref{eq:alpha_ell_relation}.
\eq{
	\frac{\ell_\mathrm{eff}^\mathrm{cNFW}}{\ell_\mathrm{fid}} = \left[ \alpha \left. \frac{\varDelta t_\mathrm{duration}^\mathrm{cNFW}}{\varDelta t_\mathrm{delay}^\mathrm{cNFW}(\phi)} \right]^{1/4} \right|_{\ell_\mathrm{eff} = \ell_\mathrm{fid}} \label{eq:alpha_ell_relation_NFW}
}
where $\ell_\mathrm{fid}$ is the fiducial effective coupling constant satisfying $\varDelta t_\mathrm{delay}^\mathrm{cNFW} = \alpha \varDelta t_\mathrm{duration}^\mathrm{cNFW}$.

From \figref{fig:signal_duration_NFW}, the signal durations for the uniform and core NFW profiles do not differ much, i.~e.~almost within a factor of three except for $r<1\,{\rm kpc}$. We note that this statement is true in all the range of axion mass that we search because the mass-dependence is the same for $\varDelta t_\mathrm{duration}^\mathrm{cNFW}$ and $\varDelta t_\mathrm{duration}$. Since the effective coupling constant for the core NFW profile, $\ell_\mathrm{eff}^\mathrm{cNFW}$, depends on $\varDelta t_\mathrm{duration}^\mathrm{cNFW}$ to the power of $1/4$ in Eq.~\eqref{eq:alpha_ell_relation_NFW}, the difference of the signal durations results in only $\sim 7\%$ shifts in the conversion of $\ell_\mathrm{eff}$ for a given $\alpha$. Thus, we assume $\varDelta t_\mathrm{duration}^\mathrm{cNFW} = \varDelta t_\mathrm{duration}$ for simplicity. By this assumption, the size of data chunks which should be used in the optimized search is equal to that in the previous search~\cite{axion_tsutsui}. That is, we do not have to search the data again and can focus on the reinterpretation of the previous constraint on the $m_a$--$\ell_\mathrm{eff}$ plane, because we did not find any axion signal under the uniform DM assumption. 

Then, to obtain a new observational constraint for the core NFW profile, we consider only the two modifications: the time delay and the amplification factor. Next we consider the modification from the time delay in the core NFW profile. From Eq.~\eqref{eq:alpha_ell_relation_NFW} with the signal duration, $\varDelta t_\mathrm{duration}$, and the time delay given in Eq.~\eqref{eq:delay_NFW}, the effective coupling, $\ell_\mathrm{eff}^\mathrm{cNFW}$, is corrected from $\ell_\mathrm{eff}^\mathrm{uni}$, which is obtained from Eq.~\eqref{eq:alpha_ell_relation} for the uniform DM profile. By defining the relative difference, $(\ell_\mathrm{eff}^\mathrm{cNFW} - \ell_\mathrm{eff}^\mathrm{uni}) / \ell_\mathrm{eff}^\mathrm{uni}$, the difference for the shift ratio $\alpha = 1$ and $m_a =  1 \times 10^{-12}\,\mathrm{eV}$ is summarized in \tabref{tab:relative_difference_bw_DMprofiles}. All the differences are negative, and the searched regions are moved to lower $\ell_\mathrm{eff}$. However, these differences are at most $\sim 11\%$ and  $\sim 3$--$8$ times smaller than the differences estimated from the modification of the amplification factor, as we will see later. Thus, we neglect the modification from the time delay together with that from the signal duration. This assumption enables us to neglect the modification on the searched region in the previous study~\cite{axion_tsutsui} and to focus simply on the modification from the amplification factor.

The $p$-value is calculated with Eq.~\eqref{eq:p_value} and equal to the threshold, $0.5\%$, on the envelope even for the both DM profiles.
Since we consider how to convert the previous search, the same values of the observed axion SNR $\chi^2_\mathrm{obs}$ are used in Eq.~\eqref{eq:p_value}.
Therefore, the different parameter to obtain the envelopes for the both DM profiles is only the noncentral parameter $\lambda$ defined in Eq.~\eqref{eq:lambda}.
That is, we can obtain the effective coupling on the envelope for the core NFW profile by solving
\eq{
	\left. F_\mathrm{total}^\mathrm{cNFW}(\phi) \right|_{\ell_\mathrm{eff} = \ell_\mathrm{eff}^\mathrm{cNFW}} = \left. F_\mathrm{total}^\mathrm{uni} \right|_{\ell_\mathrm{eff} = \ell_\mathrm{eff}^\mathrm{uni}} \label{eq:F_balance} \;,
}
with Eqs.~\eqref{eq:axion_signal_amplitude} and~\eqref{eq:lambda}.
Since the $p$-value is calculated for each GW event, the converted envelopes as the solution of Eq.~\eqref{eq:F_balance} is obtained also for each GW event.
Therefore, we have to combine the envelopes for all GW events by taking the Logical OR, as explained in \secref{sec:search_method_optimized_for_the_axion_signals}.
As there is no unrejected $\ell_\mathrm{eff}$ above the envelopes, the Logical OR is equal to taking the minimum value of the constraints on the coupling among the GW events.

\begin{figure}
	\centering
	\includegraphics[width=\linewidth]{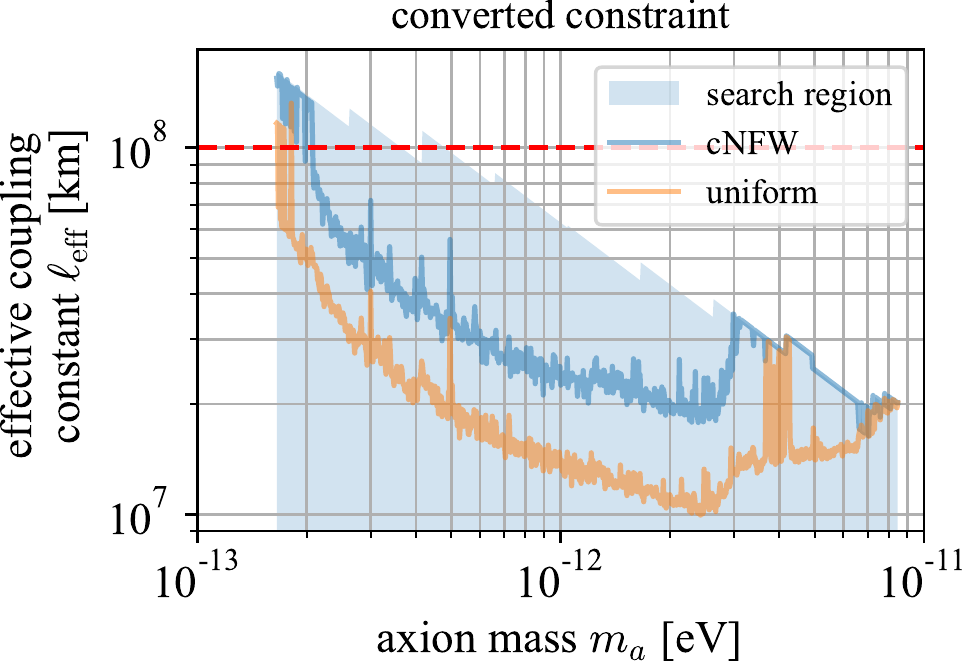}
	\caption{The combined constraint on the effective coupling constant $\ell_\mathrm{eff}$ as a function of the axion mass $m_a$ from the analysis of five GW events. The blue line is the constraints for the core NFW profile, while the orange one is the previous constraints~\cite{axion_tsutsui} for the uniform DM profile. The red dashed line is the constraint from Gravity Probe B~\cite{GravityProbeB} for $f_\mathrm{DM} = 1$. The faint blue region is the searched parameter space.}
	\label{fig:converted_combined_result}
\end{figure}

With the above method, we convert and combine the envelopes for the GW events, as shown in \figref{fig:converted_combined_result}.
If the converted results $\ell_\mathrm{eff}^\mathrm{cNFW}$ are outside the searched range, the converted constraints are replaced with the values of $\ell_\mathrm{eff}$ at the boundary of the searched range.
Since $F_\mathrm{total}^\mathrm{cNFW}$ is smaller than $F_\mathrm{total}^\mathrm{uni}$ for the GW events, the converted result becomes weaker than the previous one.
However, the converted constraint is still at most about nine times stronger than the constraint from Gravity Probe B, which is still strong enough.

So far we have considered the core NFW profile, which well explains the observations~\cite{coreNFW_profile1, coreNFW_profile2, coreNFW_profile3}.
In reality, however, there should be a true DM profile that slightly differs from the core NFW profile.
As the difference between the true and the core NFW profile must be much smaller than that between the uniform and the core NFW profile,
thus the true envelope should be closer to the converted result. That is, we conclude that the constraint is robust to the change of the DM profile.

In this paper, we use the Virial theorem to derive the velocity dispersion, $\varDelta v^\mathrm{cNFW}(r)$ in Eq.~\eqref{eq:velocity_dispersion_NFW}~\cite{DMvelocity1, DMvelocity2, DMvelocity3}. However, some readers may disagree with this derivation due to uncertainties in observations and simulations of the velocity dispersion caused by factors such as star formation, supernovae, and lower resolution simulations~\cite{review_axion}. To estimate the effect of such uncertainties on our results, we consider an extreme case where $\varDelta v^\mathrm{cNFW}(r)$ is replaced by a constant velocity of $200,\mathrm{km}$~\cite{DMvelocity1, DMvelocity2, DMvelocity3} and repeat all the calculations in this section again. The resultant constraint is only $\sim 8\%$ stronger than the one using $\varDelta v^\mathrm{cNFW}(r)$ of the core NFW in the mass range we searched. This is because the five GWs analyzed in this paper~\cite{axion_tsutsui} propagate through the edge region of the core NFW density profile, where a constant velocity of $200,\mathrm{km}$ is a good approximation. Therefore, we find that the constraint we obtained is robust to the uncertainties in the velocity dispersion.

\section{Conclusion} 
\label{sec:conclusion}
In this paper, we consider the possibility of axions being a DM candidate. The axions form coherent clouds that can delay and amplify GWs at the resonant frequency, $f_\mathrm{res}$ in Eq.~\eqref{eq:f_res}. All the GWs detected to date from compact binary mergers propagate through the Milky Way halo that consists of axion DM. Consequently, the primary GW signals are followed by secondary GW signals.

Since the properties of the secondary GWs depend on the axion mass $m_a$ and the effective coupling $\ell_\mathrm{eff}$, we can constrain $\ell_\mathrm{eff}$ by searching for the characteristic axion signals. In a previous study~\cite{axion_tsutsui}, we developed the search method optimized for the characteristic axion signals, and obtained the upper bound of $\ell_\mathrm{eff} \lesssim 1 \times 10^{7}\,\mathrm{km}$ for $m_a \in [1.7 \times 10^{-13}, 8.5 \times 10^{-13}] \,\mathrm{eV}$.  This constraint represents a ten-fold improvement over the previous constraint by Gravity Probe B~\cite{GravityProbeB}, which was $\ell_\mathrm{eff} \lesssim 1\times 10^8,\mathrm{km}$.

However, in the previous search~\cite{axion_tsutsui}, we assumed that DM is homogeneously distributed in the MW halo, which is unrealistic. Thus, in this paper, we consider the core NFW profile as a realistic DM profile. Furthermore, we consider the DM velocity dispersion estimated from the Virial theorem, $\varDelta v^\mathrm{cNFW}$ in Eq.~\eqref{eq:velocity_dispersion_NFW}. With the setting, the properties of the axion signals are modified from those for the uniform DM profile: the time delay in Eq.~\eqref{eq:delay_NFW}, the signal duration in Eq.~\eqref{eq:duration_NFW}, and the amplification factor in Eqs.~\eqref{eq:Ftotal_NFW_coherent} and~\eqref{eq:Ftotal_NFW_incoherent}. Since the finite separation between the Earth and the Galactic center is considered, the modifications above depend also on a GW propagation direction. That is, GWs through the axion clouds in the MW halo becomes anisotropic.

With such modifications of the axion signals, we consider how to convert the previous constraint for the uniform density profile~\cite{axion_tsutsui} to that for the core NFW profile. From the converted result in \figref{fig:converted_combined_result}, the constraint becomes weaker by only a factor of $\sim 2$. Thus, we conclude that the constraint is robust to the differences of the DM profiles.

\acknowledgments
T.~T. is supported by JSPS KAKENHI Grant No. 21J12046. A.~N. is supported by JSPS KAKENHI Grants No. JP19H01894 and No. JP20H04726 and by Research Grants from the Inamori Foundation.
The authors are grateful for computational resources provided by the LIGO Laboratory and supported by National Science Foundation Grants No. PHY-0757058 and No. PHY-0823459.
This material is based upon work supported by NSF's LIGO Laboratory which is a major facility fully funded by the National Science Foundation.

\appendix
\section{Jacobian from the radius to the propagating ratio of the path} \label{sec:jacobian_from_the_radius_to_the_propagating_ratio_of_the_path}

\begin{figure}[t]
	\centering
        \includegraphics[width=0.8\linewidth]{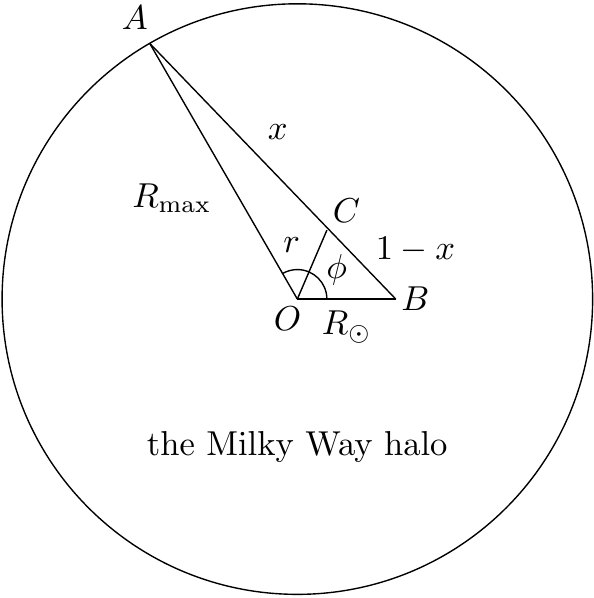}
	\caption{The system considered in \secref{sec:properties_of_secondary_gravitational_waves} and the parameters used in Appendix~\ref{sec:jacobian_from_the_radius_to_the_propagating_ratio_of_the_path}. The Galactic center is at the point $O$, the GWs enter into the MW halo from the point $A$, the Earth is at the point $B$, and the point $C$ divides $\overline{AB}$ into two segments with the ratio of $x:(1 - x)$.}
	\label{fig:system_with_parameter}
\end{figure}


In this paper, we consider the inhomogeneous DM profile (core NFW profile) and take into account the separation of the Earth from the Galactic center. Thus, some integrations depend on the paths along which GWs propagate. On the path, the radius from the Galactic center does not monotonically decrease, and then the radius is not an optimal parameter for the integrations. Then, in this Appendix, we discuss a better parameter for the integrations.

First, for convenience of an explanation, some points are named like \figref{fig:system_with_parameter}:  The Galactic center is at the point $O$, a GW enters into the MW halo from the point $A$, the Earth is at the point $B$, and the point $C$ divides the line segment $\overline{AB}$ into two segments with the ratio of $x:(1 - x)$.
That is,
\eq{
	|\overrightarrow{OA}| =& R_\mathrm{max} = 300\,\mathrm{kpc} \;, \\
	|\overrightarrow{OB}| =& R_\odot = 8.0\,\mathrm{kpc} \;, \\
	|\overrightarrow{AC}| : |\overrightarrow{BC}| =& x : 1 - x \;.
}
Thus, $x = 0$ corresponds to $\overrightarrow{OC} = \overrightarrow{OA}$ (the entrance point of the GW to the MW halo) and $x = 1$ corresponds to $\overrightarrow{OC} = \overrightarrow{OB}$ (the Earth).
Also, the parameters are defined as
\eq{
	r \coloneqq& |\overrightarrow{OC}| \;, \\
	\phi \coloneqq& \arccos \left( \frac{ \overrightarrow{OA} \cdot \overrightarrow{OB}}{ R_\mathrm{max} R_\odot } \right) \;.
}
Although an additional angle parameter is needed to express coordinates in three-dimensional space, the system defined in \figref{fig:system_with_parameter} is axisymmetric around $\overrightarrow{OB}$ because of the spherical symmetry of the core NFW profile, and needs the single angle parameter, $\phi$.
That is, without loss of generality, we can always consider the plane spanned by $\overrightarrow{OA}$ and $\overrightarrow{OB}$.

From \figref{fig:system_with_parameter}, we can write
\eq{
	\overrightarrow{OC} =& \overrightarrow{OA} + x ( \overrightarrow{OB} - \overrightarrow{OA} ) \\
		=& (1 - x) \overrightarrow{OA} + x \overrightarrow{OB} \;.
}
Then, the norm is
\eq{
	& r(x, \phi) \nonumber \\
	=& R_\mathrm{max} \sqrt{(1 - x)^2 + x^2 \left( \frac{R_\odot}{R_\mathrm{max}} \right)^2 + 2x(1 - x) \frac{R_\odot}{R_\mathrm{max}} \cos\phi} \;.
}

Next, we derive the Jacobian, $\diff r / \diff x$.
We note that $\phi$ is the GW propagation direction, that is, constant.
Thus, the Jacobian is
\eq{
	& \frac{\diff r}{\diff x}(x, \phi) \nonumber \\
	=& R_\mathrm{max} \frac{-(1 - x) + x \left( \frac{R_\odot}{R_\mathrm{max}} \right)^2 + (1 - 2x) \frac{R_\odot}{R_\mathrm{max}} \cos\phi}{\sqrt{(1 - x)^2 + x^2 \left( \frac{R_\odot}{R_\mathrm{max}} \right)^2 + 2x(1 - x) \frac{R_\odot}{R_\mathrm{max}} \cos\phi}} \;.
}
The sign of the Jacobian is flipped for some $\phi$.
However, all physical quantities discussed in \secref{sec:the_properties_of_the_secondary_gravitational_waves} are positive.
Thus, the absolute value of the Jacobian should be used in the calculation.


\bibliographystyle{unsrt}
\bibliography{references}

\end{document}